\documentclass[aps,prd,twocolumn,superscriptaddress,
nofootinbib,showpacs,balancelastpage,nobibnotes]{revtex4}
\usepackage{graphicx}
\usepackage{amsmath}

\begin{document}

\title{Sensitivity to $\theta_{13}$ and $\delta$ in the Decaying
Astrophysical Neutrino Scenario}

\author{John F. Beacom}

\author{Nicole F. Bell}
\affiliation{NASA/Fermilab Astrophysics Center, Fermi National Accelerator
Laboratory, Batavia, Illinois 60510-0500}

\author{Dan Hooper}
\affiliation{Denys Wilkinson Laboratory, Astrophysics Department,
OX1 3RH Oxford, England UK}

\author{Sandip Pakvasa}
\affiliation{Department of Physics and Astronomy, University of Hawaii,
Honolulu, Hawaii 96822}

\author{Thomas J. Weiler}
\affiliation{Department of Physics and Astronomy, Vanderbilt University,
Nashville, Tennessee 37235}

\date{23 September 2003}

\begin{abstract}
We have previously shown that the decay of high-energy neutrinos from
distant astrophysical sources would be revealed by flavor ratios that
deviate strongly from the $\phi_{\nu_e}:\phi_{\nu_\mu}:\phi_{\nu_\tau}
= 1:1:1$ expected from oscillations alone.  Here we show that the
deviations are significantly larger when the mixing angle
$\theta_{13}$ and the CP phase $\delta$ are allowed to be nonzero.  If
neutrinos decay, this could allow measurement of $\theta_{13}$ and
$\delta$ in IceCube and other near-term neutrino telescopes.
\end{abstract}

\pacs{95.85.Ry, 96.40.Tv, 14.60.Pq \hspace{4.5cm}
FERMILAB-Pub-03/292-A}


\maketitle


Traveling over cosmological distances, neutrino wave packets decohere
into mass eigenstates.  The probability to measure a neutrino flavor
$\beta$ at Earth is therefore
\begin{equation}
\label{eq:one}
P_\beta = \sum_\alpha  w_\alpha\,
\sum_j |U_{\alpha j}|^2 \,|U_{\beta j}|^2 \,,
\end{equation}
where $U_{\alpha j}$ are elements of the neutrino mixing matrix, and
$w_\alpha$ are the weights of the flavors produced in the
astrophysical source.  As is well-known, the weights from a pion-muon
decay chain, $w_e:w_\mu:w_\tau = 1:2:0$, lead to flavor ratios at
Earth of $\phi_{\nu_e}:\phi_{\nu_\mu}:\phi_{\nu_\tau} =
1:1:1$~\cite{111}.  Variation of the mixing angles from the assumed
$\nu_\mu - \nu_\tau$ symmetry limit ($\theta_{23}=45^\circ$ and
$\theta_{13}=0$) leads to only small ($\lesssim 20\%$) deviations.

We have recently shown that the expected flavor ratios would be
dramatically altered if neutrinos decay~\cite{cosmicDK}.  The
strongest lifetime limits, from solar neutrinos, are too weak to
restrict the possibility of astrophysical neutrino decay by a factor
of about $10^7$~\cite{solarDK,cosmicDK}.  Other scenarios,
small-$\delta m^2$ active-sterile mixing in pseudo-Dirac~\cite{pDirac}
or mirror models~\cite{steriles}, or CPT violation~\cite{Barenboim},
produce more subtle deviations from the expected $1:1:1$.  While
neutrino decay can be tested with IceCube~\cite{IceCube} and other
near-term detectors, the latter scenarios may require future
detectors~\cite{flavors}.

The most interesting decay scenario is that in which $\nu_3$ and
$\nu_2$ decay (either into active $\nu_1$ or sterile states), but the
lightest neutrino $\nu_1$ is stable.  In this case, the beam contains
just the single mass eigenstate $\nu_1$, and the flavor ratios at
Earth are simply
\begin{equation}
\label{eq:two}
\phi_e:\phi_\mu:\phi_\tau = |U_{e1}|^2:|U_{\mu 1}|^2:|U_{\tau 1}|^2\,.
\end{equation}
We present here further analysis of this scenario, taking into account
the broken $\nu_\mu - \nu_\tau$ symmetry that arises when $U_{e3} =
\sin\theta_{13} e^{-i\delta} \neq 0$.  Even for small $\theta_{13}$, we
find that the flavor ratios are very sensitive to the CP phase
$\delta$, which was set to zero for simplicity in our earlier
work~\cite{cosmicDK}.

We assume a normal neutrino mass hierarchy.  For the case of an
inverted hierarchy (or no decay), varying $\theta_{13}$ has little
effect.  Note that in this decay scenario, there is no dependence on
the initial astrophysical flux ratios, and hence on the production
mechanism of the ultra high energy neutrinos, since the flux reaching
Earth consists only of the lightest neutrino $\nu_1$.

\begin{figure}
\includegraphics[width=3.25in]{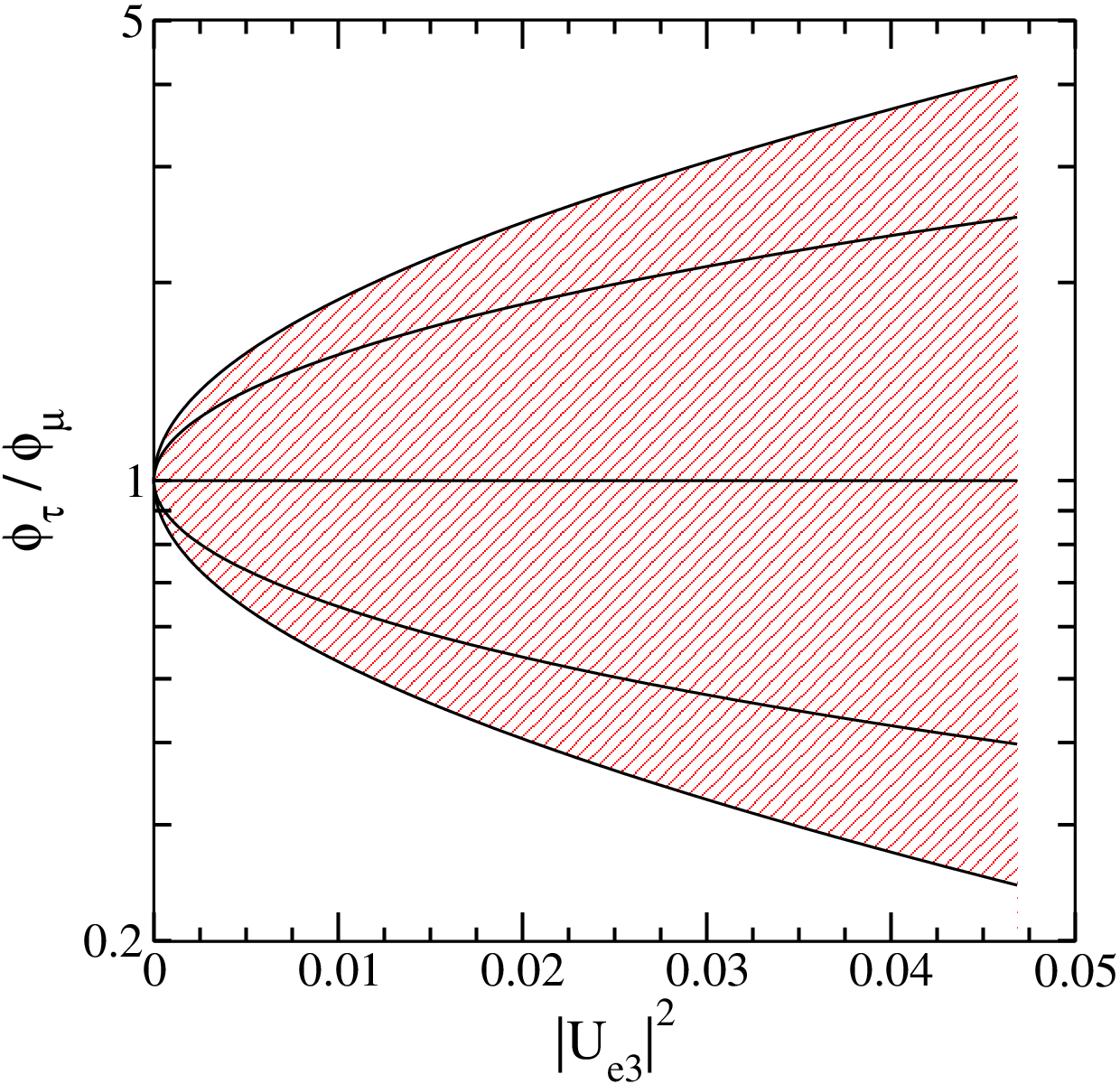}
\caption{\label{fig:taumu} Variation of the ratio $\phi_\tau/\phi_\mu$
with $|U_{e3}|^2 = s_{13}^2$ in the allowed range~\cite{reactors}.
From bottom to top, the solid curves correspond to $\delta = (0,
\pi/4, \pi/2, 3\pi/4, \pi)$, with the hatched region showing the full
allowed range.  The atmospheric angle is $\theta_{23} =
45^\circ$~\cite{atm} and the solar angle is $\theta_{12} =
32.5^\circ$~\cite{solar}.}
\end{figure}

Expressing the flavor ratios in Eq.~(\ref{eq:two}) in terms of the
mixing parameters (using the conventions of Ref.~\cite{RPP} and
$c_{ij} \equiv \cos \theta_{ij}$, $s_{ij} \equiv \sin \theta_{ij}$),
\begin{equation}
\label{taumu}
\frac{\phi_\tau}{\phi_\mu} =
\frac{s_{12}^2 (1 - \cos2\theta_{23}) +
s_{13}^2 c_{12}^2 (1 + \cos2\theta_{23}) - 4\hat{J}}
{s_{12}^2 (1 + \cos 2\theta_{23}) +
s_{13}^2 c_{12}^2 (1 - \cos2\theta_{23}) + 4\hat{J}}\,.
\end{equation}
We have defined
\begin{equation}
\label{jhat}
\hat{J} \equiv
\frac{1}{4}\sin2\theta_{12}\,\sin2\theta_{23}\,(s_{13} \cos\delta)\,,
\end{equation}
which is related to the Jarlskog invariant $J$ according to $J =
\hat{J} c_{13}^2 \tan \delta$.  It is known, and is evident in
Eq.~(\ref{taumu}), that either a nonzero $\cos 2\theta_{23}$ or
$s_{13}$ breaks $\nu_\mu - \nu_\tau$ symmetry.  In
Fig.~\ref{fig:taumu}, we show the ratio $\phi_{\tau}/\phi_{\mu}$ as a
function of $|U_{e3}|^2$ (within the range allowed by reactor
experiments~\cite{reactors}), with $\theta_{23}= \pi/4$ fixed.  Even a
small $\theta_{13}$ has a relatively large effect, particularly when
the CP phase $\delta$ is allowed to be nonzero.  However, direct
measurement of the ratio $\phi_\tau/\phi_\mu$ is very difficult, since
events which are unique to $\nu_\tau$ (double-bang and lollipop
events) have much lower detection probabilities~\cite{flavors}.

\begin{figure}
\includegraphics[width=3.25in]{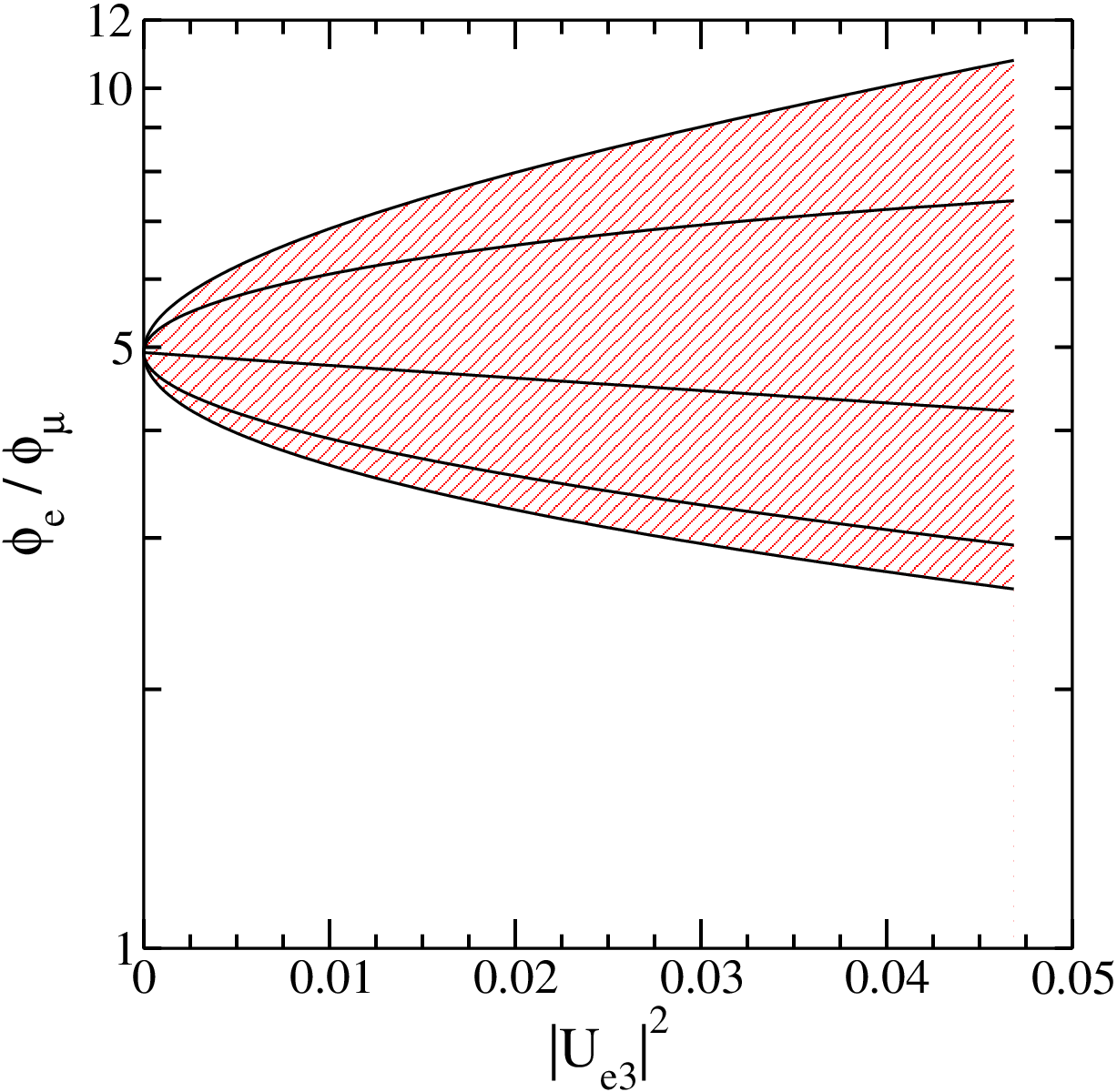}
\caption{\label{fig:emu} Same as Fig.~\ref{fig:taumu}, except for the
ratio $\phi_e/\phi_\mu$.  As in Fig.~\ref{fig:taumu}, the value
expected in the no-decay case is 1; here the effects of decay are
always pronounced.}
\end{figure}

In contrast, the $\phi_e/\phi_\mu$ ratio can be directly probed in a
detector like IceCube by comparing the rate of shower events to muon
events~\cite{flavors}.  This flavor ratio is
\begin{equation}
\label{emu}
\frac{\phi_e}{\phi_\mu} =
\left(\frac{|U_{e1}|^2}{1 - |U_{e1}|^2}\right) 
\left(1 + \frac{\phi_\tau}{\phi_\mu}\right)\,.
\end{equation}
The $\nu_e$ fraction in the $\nu_1$ mass eigenstate is insensitive to
values of $\theta_{13}$ in the allowed range, shown by the first
factor in Eq.~(\ref{emu}), where $|U_{e1}| \equiv (c_{12}c_{13})^2$.
However, the broken $\nu_\mu - \nu_\tau$ symmetry affects the
$\phi_e/\phi_{\mu}$ ratio through the second factor in
Eq.~(\ref{emu}), and this is a large effect.  When $\cos \delta$ is
negative (positive) it decreases (increases) the $\nu_\mu$ fraction of
$\nu_1$ with respect to the $\nu_\tau$ fraction, resulting in an
enhanced (suppressed) $\nu_e/\nu_\mu$ ratio.  This is shown in
Fig.~\ref{fig:emu}.  The curve with $\delta = 0$ is as in
Ref.~\cite{cosmicDK}, though we have updated the solar angle
$\theta_{12}$.  For nonzero $\delta$, new to this work, the flavor
ratio is significantly farther from the no-decay value of 1.

Note that the dependence of Eqs.~(\ref{taumu}) and (\ref{emu}) on the
CP phase $\delta$ occurs only through $\cos\delta$, since this is a
CP-conserving observable.  Therefore, it is not necessary to separate
astrophysical neutrinos and antineutrinos, which would be very
difficult.  The phase $\delta$ is of crucial importance in terrestrial
long-baseline oscillation experiments since CP-violating observables,
based on the comparison of neutrino and antineutrino oscillation
probabilities, are proportional to $\sin\delta$~\cite{LB}.  Farzan and
Smirnov have shown that a nonzero $\sin\delta$ may in principle also
be inferred by direct construction of the leptonic unitarity
triangle~\cite{Farzan}.  A key distinction is that the terrestrial
experiments use a beam of flavor eigenstates, whereas neutrino decay
can produce a pure mass eigenstate, allowing for very large variation
with $\delta$.  Since measurement of $\delta$ in terrestrial
experiments will be an extremely challenging task, it is intriguing to
find an example where the effect of varying $\delta$ is huge.
Finally, the flavor ratios are sensitive only to the ``Dirac'' phase
$\delta$, and not the ``Majorana'' phases; Majorana phases are
relative phases between mass eigenstates, and the beam consists of the
single mass eigenstate $\nu_1$.

\begin{figure}
\includegraphics[width=3.25in]{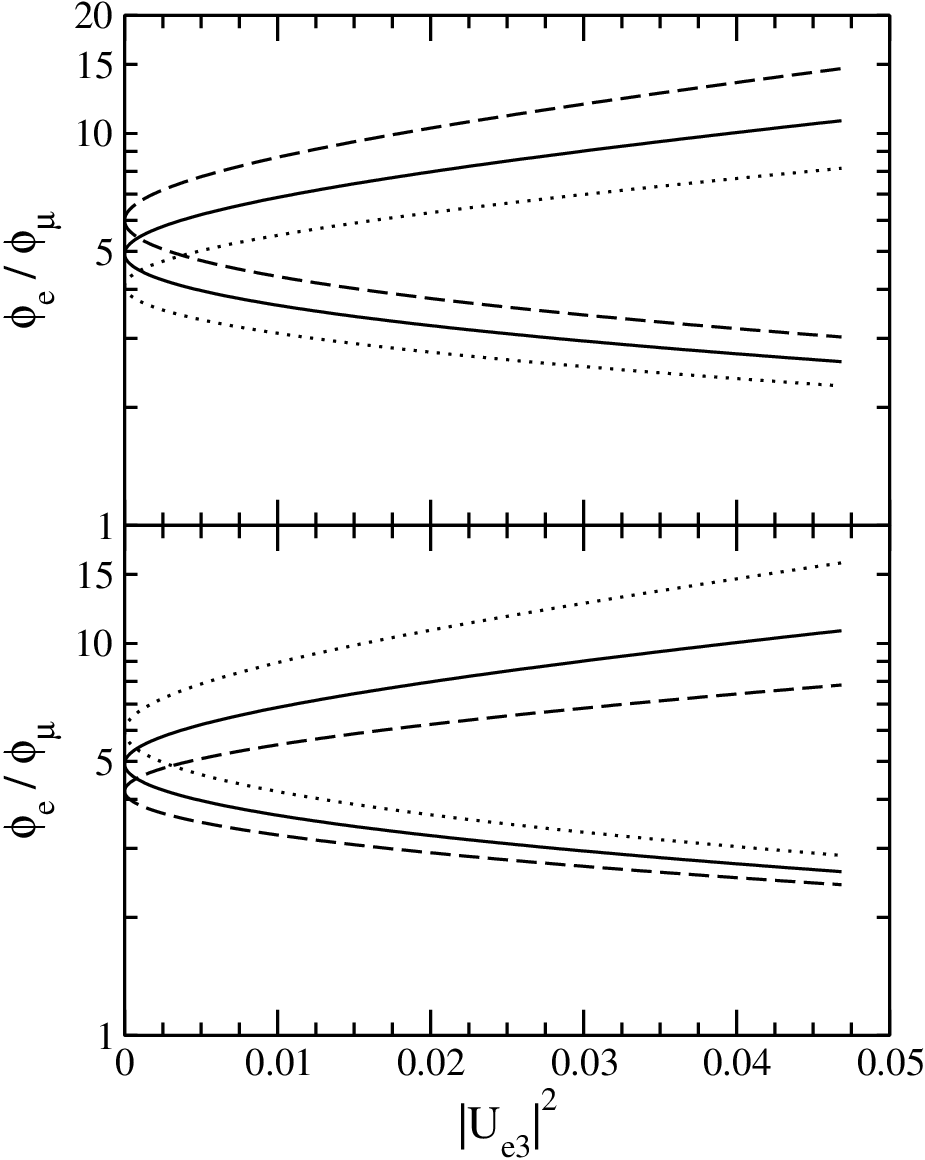}
\caption{\label{fig:emuvar} Upper panel: varying the solar angle, with
$\theta_{12} = 30^\circ$ (dashed), $32.5^\circ$ (solid) and $35^\circ$
(dotted).  Lower panel: varying the atmospheric angle, with
$\theta_{23} = 40^\circ$ (dashed), $45^\circ$ (solid) and $50^\circ$
(dotted).  The bottom and top curves correspond to $\delta = (0,
\pi)$.  As before, the region between the curves is the allowed range,
obtained for different values of $\delta$.}
\end{figure}

Variation of the atmospheric mixing angle $\theta_{23}$ away from
$45^\circ$ also breaks the $\nu_\mu - \nu_\tau$ symmetry, as shown in
Eq.~(\ref{taumu}), and has a similar effect on the flavor ratios.
Variation of the solar mixing angle changes the $\phi_e/\phi_\mu$
ratio as it alters the $\nu_e$ fraction of $\nu_1$.  In
Fig.~\ref{fig:emuvar}, we show how the variations of these angles
within their one-sigma allowed ranges affects $\phi_e/\phi_\mu$.  Note
that the size of the variation due to uncertainties in the solar and
atmospheric angles are quite similar.  These uncertainties will be
reduced by existing or planned solar and long-baseline experiments.
The angle $\theta_{13}$ may be measured by future long
baseline~\cite{LB} or reactor~\cite{reactor13} experiments, and
measuring $\delta$ may require a neutrino factory.

To conclude, IceCube and other detectors have an excellent chance of
detecting astrophysical neutrinos and measuring their flavor ratios in
the next several years.  If neutrinos decay, the flavor ratio
$\phi_e/\phi_\mu$ will be much larger than its no-decay value of 1,
and this effect is significantly enhanced by nonzero $\theta_{13}$ and
$\delta$.  Thus there may be a new opportunity to measure the last
unknown values in the neutrino mixing matrix.


{\bf Acknowledgments.---}
We thank Stephen Parke for illuminating discussions.  J.F.B. and
N.F.B. were supported by Fermilab (operated by URA under DOE contract
DE-AC02-76CH03000) and NASA grant NAG5-10842; S.P. by DOE grant
DE-FG03-94ER40833; and T.J.W. by DOE grant DE-FG05-85ER40226, NASA
grant NAG5-13399, and a Vanderbilt Discovery Award.  T.J.W. also
thanks the U. Wisconsin Phenomenology Group for hospitality.


\end{document}